\documentclass[times]{sigplanconf}

\usepackage{amsmath}
\usepackage{url}
\usepackage{graphicx}
\usepackage{subfig}
\usepackage{algorithm}
\usepackage{algorithmic}
\usepackage{tabularx}
\usepackage{amsmath}
\usepackage{multirow}
\usepackage[colorlinks,
            linkcolor=blue,
            anchorcolor=blue,
            citecolor=blue,
            urlcolor=blue,
            bookmarksdepth=3]{hyperref}

\begin{document}

\copyrightyear{2015}
\copyrightdata{[to be supplied]}


\title{On the Shoulders of Giants: Incremental Influence Maximization in Evolving Social Networks}

\authorinfo{Xiaodong Liu\and Xiangke Liao\and Shanshan Li\and Jingying Zhang\and Lisong Shao\and Chenlin Huang\and Liquan Xiao}
           {National University of Defense Technology, Changsha, 410073, China}
           {\{liuxiaodong, xkliao, shanshanli, jyzhang, lisongshao, chenlinhuang, lqxiao\}@nudt.edu.cn}

\maketitle

\begin{abstract}
Identifying the most influential individuals can provide invaluable help in developing and deploying effective viral marketing strategies. Previous studies mainly focus on designing efficient algorithms or heuristics to find top-$K$ influential nodes on a given static social network. While, as a matter of fact, real-world social networks keep evolving over time and a recalculation upon the changed network inevitably leads to a long running time, significantly affecting the efficiency.
In this paper, we observe from real-world traces that the evolution of social network follows the preferential attachment rule and the influential nodes are mainly selected from high-degree nodes. Such observations shed light on the design of IncInf, an incremental approach that can efficiently locate the top-$K$ influential individuals in evolving social networks based on previous information instead of calculation from scratch.
In particular, IncInf quantitatively analyzes the influence spread changes of nodes by localizing the impact of topology evolution to only local regions, and a pruning strategy is further proposed to effectively narrow the search space into nodes experiencing major increases or with high degrees.
We carried out extensive experiments on real-world dynamic social networks including Facebook, NetHEPT, and Flickr. Experimental results demonstrate that, compared with the state-of-the-art static heuristic, IncInf achieves as much as $21\times$ speedup in execution time while maintaining matching performance in terms of influence spread.
\end{abstract}

\category{G.2.2}{Graph Theory}{Graph algorithms and network problems}

\terms
Algorithms, Performance

\keywords
Influence maximization, incremental algorithm, evolving social network, graph algorithm

\section{Introduction}
\label{intro}
Influence maximization (IM) is one fundamental and important problem which aims to identify a small set of influential individuals so as to develop effective viral marketing strategies in large-scale social networks\cite{domin2001}. As a matter of fact, real-world social networks keep evolving over time. For example, in Facebook, new people might join while old ones might withdraw, and people might make new friends with each other. Moreover, real-world social networks are evolving in a rather surprising speed; it is reported that as much as 1 million new accounts are created in Twitter every day \cite{twitter}. Such massive evolution of network topology, on the contrary, may lead to a significant transformation of the network structure, thus raising a natural need of efficient reidentification.

Existing researches and solutions on influence maximization focus mainly on developing effective and efficient algorithms on a given static social network. Although one could possibly run any of the static influence maximization methods, such as \cite{chen09,chen10,jure07,esmce}, to find the new top-$K$ influential individuals when the network is updated, this approach has some inherent drawbacks that cannot be neglected: (1) the running time of a specific static method can be extremely long and unacceptable especially on large-scale social networks, and (2) whenever the network topology is changed, we need to recalculate the influence spreads for all the nodes which leads to very high costs. Can we quickly and efficiently identify the influential nodes in evolving social networks? Can we incrementally update the influential nodes based on previously known information instead of frequently recalculating from scratch?

Unfortunately, the rapidly and unpredictably changing topology of a dynamic social network poses several challenges in the reidentification of influential users, which we list as follows. On one hand, the interconnections between edges in real-world social graphs are rather complicated; as a result, even one small change in topology may affect the influence spreads of a large number of nodes, not to mention the massive changes in large-scale social networks. It is very difficult to efficiently compute the changes of influence spreads for all the nodes after the evolution. On the other hand, since there are a great number of nodes in large-scale social networks, how to effectively limit the range of potential influential nodes and reduce the amount of calculation to the maximum is a very challenging problem.

To well address these challenges, we investigate the dynamic characteristics exhibited during the evolution of real-world social networks. Through tests on three real-world dataset traces, Facebook, NetHEPT and Flickr, we observe that, first, the growth of social network is mainly based on the preferential attachment principle \cite{pa1999}, that is the new-coming edges prefer to attach to nodes with higher degree, which naturally leads to the ``rich-get-richer" phenomena; and second, the top-$K$ influential nodes are mainly selected from those high-degree nodes. Inspired by such observations, we know that the influence changes of some nodes will have no impact on the top-$K$ selection, and thus can be pruned to reduce the amount of calculation. Motivated by this, we propose IncInf, an incremental method to identify the top-$K$ influential nodes in evolving social networks instead of recalculating from scratch, thus significantly improving the efficiency and scalability to handle extraordinarily large-scale networks. To summarize, the main contributions of IncInf are as follows:

First, we design an efficient approach to quantitatively analyze the influence spread changes from network topology evolution by adopting the idea of localization. A tunable parameter is provided to tradeoff between efficiency and effectiveness.

Second, we propose a pruning strategy which could effectively narrow the search space into nodes only experiencing major increases or with high degrees based on the changes of influence spread and the previous top-$K$ information.

Third, we conduct extensive experiments on three dynamic real-world social networks. Compared with the state-of-the-art static algorithm, IncInf achieves up to $21\times$ speedup in execution time while providing matching influence spread. Moreover, IncInf provides better scalability to scale up to extraordinarily large-scale networks.

The remainder of this paper is organized as follows. Section \ref{sec:2} presents related preliminaries and problem definition. Section \ref{sec:3} shows the structural evolution characteristics of dynamic social networks that we observe from three datasets: Facebook, NetHEPT and Flickr. Section \ref{sec:4} details the design of our incremental algorithm IncInf. The performance of IncInf is evaluated by comprehensive experiments in Section \ref{sec:5}. We present related work in Section \ref{sec:6} and conclude in Section \ref{sec:7}.

\section{Preliminaries and Problem Statement}
\label{sec:2}
In this section, we illustrate the definition of social network and the influence diffusion model that we will use throughout the paper, and then give the problem definition of influence maximization in evolving networks.
\subsection{Preliminaries on Influence Maximization}
\label{sec:21}
\paragraph{Social Network.} A social network is formally defined as a directed graph $G=(V,E,P)$ where node set $V=\{v_1,v_2,\cdots,v_n\}$ denotes entities in the social network. Each node can be either active or inactive, and will switch from being inactive to being active if it is influenced by others nodes. Edge set $E \subset V\times V$ is a set of directed edges representing the relationship between different users. Take Twitter as an example. A directed edge $(v_i,v_j)$ will be established from node $v_i$ to $v_j$ if $v_i$ is followed by $v_j$, which indicates that $v_j$ may be influenced by $v_i$. $P$ denotes the influence probability of edges; each edge $(v_i,v_j)\in E$ is associated with an influence probability $p({v_i},{v_j})$ defined by function $p:E \to [0,1]$. If $\left( {{v_i},{\rm{ }}{v_j}} \right) \notin E$, then $p({v_i},{v_j}) = 0$.

\begin{algorithm}
\caption{Basic Greedy}
\label{alg:basicgreedy}
\begin{algorithmic}[1]
\STATE Initialize $S=\emptyset $\label{alg:basicgreedy:1}
\FOR{$i=1$ to $K$}\label{alg:basicgreedy:2}
    \STATE Select $v = \arg\max _ {v_i\in (V\backslash S)} (\sigma(S\cup v_i)-\sigma(S))$\label{alg:basicgreedy:3}
    \STATE $S = S \cup \{v\}$\label{alg:basicgreedy:4}
\ENDFOR\label{alg:basicgreedy:5}
\end{algorithmic}
\end{algorithm}
\paragraph{Independent Cascade (IC) Model.} IC model is a popular diffusion model that has been well-studied in \cite{chen09,sa,kempe2003,esmce,cga}. Given an initial set $S$, the diffusion process of IC model works as follows. At step 0, only nodes in $S$ are active, while other nodes stay in the inactive state. At step $t$, for each node $v_i$ which has just switched from being inactive to being active, it has a single chance to activate each currently inactive neighbor $v_j$, and succeeds with a probability $p({v_i},{v_j})$. If $v_i$ succeeds, $v_j$ will become active at step $t+1$. If $v_j$ has multiple newly activated neighbors, their attempts in activating $v_j$ are sequenced in an arbitrary order. Such a process runs until no more activations are possible \cite{kempe2003}. We use $\sigma (S)$ to denote the influence spread of the initial set $S$, which is defined as the expected number of active nodes at the end of influence propagation.

\paragraph{Basic Greedy Algorithm.} Richardson and Domingos \cite{domin2001,domin2012} first introduced the influence maximization problem on static networks in 2001. In \cite{kempe2003}, Kempe et al. propose a basic hill-climbing greedy algorithm as shown in Algorithm \ref{alg:basicgreedy}. The proposed greedy algorithm works in $K$ iterations, starting with an empty set $S$ (line \ref{alg:basicgreedy:1}). In each iteration, a node $v_i$ which brings the maximum marginal influence spread $\sigma_S {(v_i)}=\sigma(S\cup v_i )-\sigma(S)$ is selected to be included in $S$ (lines \ref{alg:basicgreedy:3} and \ref{alg:basicgreedy:4}). The process ends when the size of $S$ reaches $K$ (line \ref{alg:basicgreedy:2}). However, this algorithm has a serious efficiency drawback due to the compute-intensive influence spread calculation. Several recent studies \cite{chen09,chen10,goyal2012,sa,irie,jure07,esmce,cga} aimed at addressing this efficiency issue.
\subsection{Formal Definition of IM problem in Evolving Networks}
\label{sec:22}
This paper differentiates itself from previous works by considering the dynamic nature of online social networks. As a matter of fact, the real-world social networks are not wholly static but keep evolving gradually over time. The evolution of large social networks has raised new sets of questions; among them one interesting yet challenging problem is how to quickly identify the top-$K$ influential users when the topology of the network is changed.

To solve such a problem, we define an evolving network $\zeta=(G^0,G^1,\cdots,G^t)$ as a sequence of network snapshots evolving over time, where ${G^t}=(V^t,E^t,P^t)$ is the network snapshot at time $t$. $\Delta{G^t}=(\Delta{V^t},\Delta{E^t},\Delta{P^t})$ denotes the structural change of network graph ${G^t}$. Obviously, we have ${G^{t+1}}={G^t}\bigcup\Delta{G^t}$. And the influence maximization problem is defined as follows:

\paragraph{\textbf{Given:}} The social network ${G^t}$ at time $t$, the top-$K$ influential nodes ${S^{t}}$ in ${G^t}$, and the structural evolution $\Delta{G^t}$ of graph ${G^t}$.
\paragraph{\textbf{Objective:}} To identify the influential nodes ${S^{t + 1}} \subset {V^{t + 1}}$ of size $K$ in ${G^{t+1}}$ at time $t+1$, such that the influence spread $\sigma (S^{t + 1})$ is maximized at the end of influence diffusion.

\section{Observations of Social Network Evolution}
In this section, we study some patterns of social network evolution. The number of nodes and edges are firstly investigated in Section \ref{sec:31} to examine the growth of users and interconnections over time. Then, we look into the degree distribution of nodes and the preferential attachment rule for new edges in Section \ref{sec:32}. We further examine the relation between the influence and the degree of node in Section \ref{sec:33}.
We study three network traces: Facebook, NetHEPT and Flickr whose detailed description can be found in Section \ref{sec:5}. Here we only show the results on Facebook since the evolution trends on the other datasets are qualitatively similar and thus omitted.
\label{sec:3}
\subsection{How Fast does the Network Evolve?}
\label{sec:31}
\begin{figure}
\centering
  \includegraphics[width=85mm]{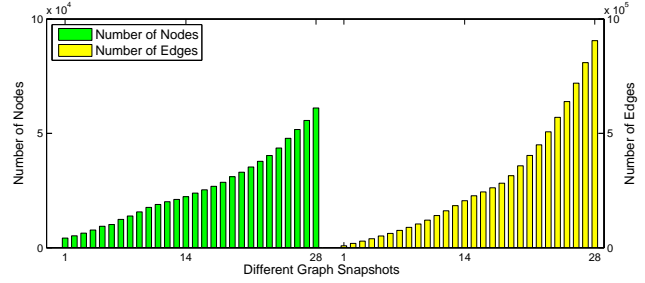}
\caption{Number of Nodes and Edges per month of the Facebook dataset.}
\label{fig:1}       
\end{figure}
Nodes and edges are the basic elements of the social network topology. In this subsection, we use the number of nodes and edges to examine the growth of users and interconnections over time. Figure \ref{fig:1} illustrates the number of nodes and edges over the entire trace period on the Facebook dataset; we take a snapshot per month. From Figure \ref{fig:1}, we observe a linear increase in the number of nodes which indicates a steady number of new users joined the network per month. While in terms of edges, the number goes up almost exponentially. The number of edges after 14 months is $25.6\times$ of that in the initial graph while the number rises to $112.9\times$ after 28 months. Such rapid growth of nodes and edges raise a natural need to efficiently find the most influential nodes after the topology evolution.
\subsection{What is the Pattern of Network Topology Evolution?}
\label{sec:32}
\begin{figure}[!t]
\centering
\subfloat[Degree distribution]{
\includegraphics[width=60mm]{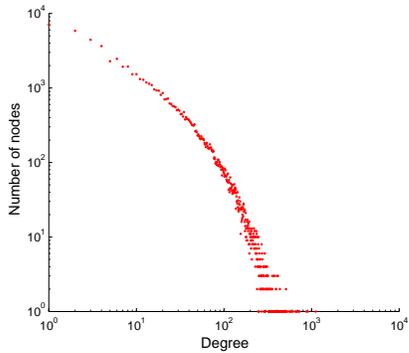}
\label{fig:2a}
}\\
\subfloat[Preferential attachment]{
\includegraphics[width=60mm]{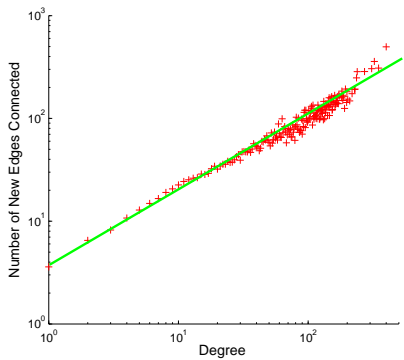}
\label{fig:2b}
}
\caption{Degree distribution and preferential attachment on Facebook.}
\label{fig:2}
\end{figure}
Understanding the pattern of the network topology evolution is of primary importance to design efficient influence maximization algorithms for evolving social networks. In this subsection, we further investigate the degree distribution of nodes and the preferential attachment rule \cite{pa1999,pa2006,pajure2008} for new coming edges. Figure \ref{fig:2a} shows the degree distribution of the Facebook final graph in log-log scale. As expected, it mainly follows the well-known power-law distribution. A large percent of the users have only a small number of links with other users, while there exist some ``hub" nodes with extremely large number of connections. This is consistent with the real-world networks.

We also study the preferential attachment rule, or in other words, the ``rich-get-richer" rule \cite{rich2010,kumar2006}, which postulates that when a new node joins the network, it creates a number of edges, where the destination node of each edge is chosen proportional to the destination's degree. This means that new edges are more likely to connect to nodes with high degree than ones with low degree. This is reasonable in reality; Lady Gaga gains 30,000 new followers on average every day \cite{ladygaga} which can never image for any common individual. The results on the Facebook dataset are demonstrated in Figure \ref{fig:2b} where the x axis is the degree of different nodes and the y axis is the average number of new edges attached to nodes of different degree. Note that both the x and y axis are in log scale. From Figure \ref{fig:2b} we can see that the degree of users in Facebook is linearly correlated with the number of new links created. This suggests that high-degree nodes get super-preferential treatment. Consequently, the influence spread change should be considerably great for the influential nodes, while there may be only small or even no change for ordinary people.

\subsection{What is the Relation between Influence and Degree?}
\label{sec:33}
\begin{figure}
  \includegraphics[width=85mm]{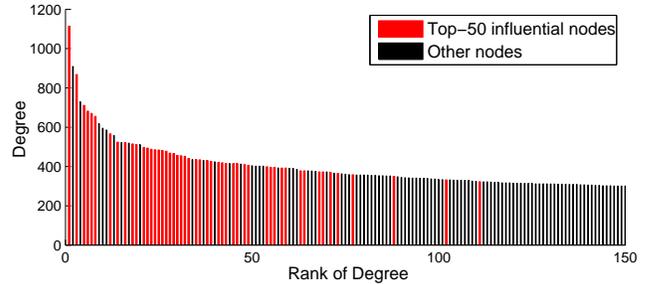}
\caption{The relation between the influence spread and the degree in Facebook.}
\label{fig:4}       
\end{figure}

Examining the relation between the influence and the degree of node can help us understand the effect of degree changing on the influence spread of nodes. For this reason, we run the static MixGreedy algorithm \cite{chen09} on the final graph and identify the top-50 influential nodes. The results on the Facebook dataset are illustrated in Figure \ref{fig:4} where the x axis is the rank of degrees of different nodes (we only show the top 150). Obviously, all the selected influential nodes have a large degree. In particular, among the 50 nodes, 48 nodes rank in top 100 of the whole 61,096 nodes in terms of degree, and the other two nodes rank 102 and 111 respectively. While on the NetHEPT and Flickr datasets, the top-50 influential nodes are selected from the top 1.79\% and 0.84\% nodes in degree, respectively. This demonstrates that the top-$K$ influential nodes are mainly selected from those with large degrees. However, it is worthy of note that the top-$K$ influential nodes in influence maximization are usually not the top-$K$ nodes ranking in degree, since the influence spread of different nodes may overlap with each other.

\section{IncInf Design}
\label{sec:4}
\begin{figure}
\centering
  \includegraphics[width=40mm]{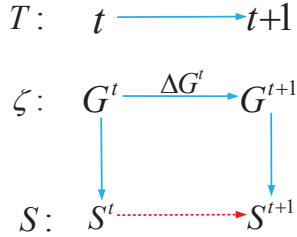}
\caption{IncInf Design.}
\label{fig:5}       
\end{figure}

\begin{table*}
\centering
\caption{Details of six types of basic operation}
\label{tab:2}       
\renewcommand{\tabularxcolumn}[1]{m{#1}}
\begin{tabularx}{170mm}{c|X|X}
\hline
\textbf{Operation} & \textbf{Description} & \textbf{Impact on influence spread}\\ \hline\hline
$addNode(u)$ & add a new node $u$ into the current network & the influence spread of $u$ is set to 1 \\ \hline
$removeNode(u)$ & delete an existing node $u$ from the network & the influence spread of $u$ is set to 0 \\ \hline
$addEdge(u,v,w)$ & introduce a new edge $\left( {u,v} \right)$ with $p\left( {u,v} \right)=w$& the influence spread of all the nodes that can reach $u$ may be increased  \\ \hline
$removeEdge(u,v)$ & remove an existing edge $\left( {u,v} \right)$ from the network & the influence spread of all the nodes that can reach $u$ may be decreased \\ \hline
$addWeight(u,v,\Delta{w})$ & increase $p(u,v)$ by $\Delta{w}$ & the influence spread of all the nodes that can reach $u$ may be increased  \\ \hline
$decWeight(u,v,\Delta{w})$ & reduce $p(u,v)$ by $\Delta{w}$ & the influence spread of all the nodes that can reach $u$ may be decreased  \\ \hline
\end{tabularx}
\end{table*}
In this section, we present the detailed design of IncInf, an  incremental approach to solve the influence maximization problem on dynamic social networks.
The main idea of IncInf is to take full use of the valuable information that is inherent in the network structural evolution and previous influential nodes, so as to substantially narrow the search space of influential nodes. In this way IncInf can significantly reduce the computation complexity and improve the efficiency. Figure \ref{fig:5} briefly illustrates the general idea of IncInf in dynamic social networks. The top-$K$ influential nodes ${S^{t + 1}}$ of ${G^{t + 1}}$ at time $t+1$ is incrementally identified based on the previous influential nodes ${S^{t}}$ at time $t$ and the structural change $\Delta{G^t}$ from ${G^{t}}$ to ${G^{t + 1}}$.
In particular, we design an efficient method to quantitatively analyze the impact of different structural changes on the influence spread of nodes by adopting the idea of localization (Section \ref{sec:42}), and propose a pruning strategy to reduce the number of potential influential nodes (Section \ref{sec:43}). We first describe six types of basic operation of topology evolution in dynamic networks in Section \ref{sec:41}.

\subsection{Basic operations of Topology Evolution}
\label{sec:41}

The evolution of social network, when reflected into its underlying graph, can be summarized into six categories, which are inserting or removing a node, introducing or deleting an edge, and increasing or decreasing the influence probability of an edge. We denote the six types of topology change as $addNode$, $removeNode$, $addEdge$, $removeEdge$, $addWeight$, $decWeight$. The detailed descriptions and their effects on influence spread are shown in Table \ref{tab:2}.

It should be noted that only after the $addNode$ operation can node $u$ establish links ($addEdge$) or sever links ($removeEdge$) with other nodes, and node $u$ can only be removed when all its associated edges are deleted. Moreover, the weight operation can be equivalently decomposed into two edge operations. For example, $addWeight(u,v,\Delta{w})$ can be divided into $removeEdge(u,v)$ and $addEdge(u,v,w+\Delta{w})$, supposing the previous weight of edge $\left( {u,v} \right)$ is $w$.

\subsection{Influence Spread Changes}
\label{sec:42}
As discussed above, whenever an edge $\left( {u,v} \right)$ is introduced into or removed from the social network, the influence spread of all the nodes that can reach node $u$ may be changed. However, as a matter of fact, the real-world social networks exhibit small-world network characteristics and the connections between nodes are highly complicated. As a result, even one small change in topology, such as an edge addition or removal, may affect the influence spread of a large number of nodes, thus introducing massive recalculations.
In order to reduce the amount of computation, we design an approach to efficiently calculate the changes on the influence spread of nodes which adopts the localization idea \cite{chen10} and tries to restrict the influence spread to the local regions of nodes.

The main idea of localization is to use the local region of each node to approximate its overall influence spread. In particular, we use the maximum influence path to approximate the influence spread from node $u$ to $v$. Here the maximum influence path $MIP(u,v,G)$ from node $u$ to $v$ in graph $G$ is defined as the path with the maximum influence probability among all the paths from node $u$ to $v$, and can be formally described as follows:
\begin{equation}\label{equ:1}
MIP(u,v,G)=\arg\max _ {p\in P(u,v,G)}\{prob(p)\}
\end{equation}
where $prob(p)$ denotes the propagation probability of path $p$ and $P(u,v,G)$ denotes all the paths from node $u$ to $v$ in graph $G$.
For a given path $p=\{u_1,u_2,\cdots,u_m\}$, the propagation probability of path $p$ is defined as follows:
\begin{equation}\label{equ:2}
prob(p)=\prod_{i=1}^{m-1}p(u_{i},u_{i+1})
\end{equation}
Moreover, an influence threshold $\theta$ is set to tradeoff between accuracy and efficiency. During the propagation process, we only consider paths whose influence probability are larger than $\theta$ while ignoring those with probability smaller than $\theta$. By doing this, the influence is effectively restricted to the local region of each node.

Similarly, in our proposal we localize the impact of topology changes on influence spread into local regions, and thus reduce the amount of computation.
Among six types of topology change, $addNode$ (or $removeNode$) is the most straightforward since it simply sets the influence spread of the node to 1 (or 0); $addWeight$, $decWeight$ as well as $removeEdge$ are methodologically similar to $addEdge$. Consequently, in the following we take $addEdge$ as an example to show which nodes' influence spread need to be updated and how to determine those changes when a new edge is added into the graph.

\begin{algorithm}
\caption{Edge addition}
\label{alg:2}
\begin{algorithmic}[1]
\REQUIRE ~~\
a new edge $e=(u,v,w)$, graph ${G^t}$.
\ENSURE ~~\
The influence spread changes of nodes in ${G^{t'}}$.
\IF {$w < \theta$ or $w\leq prob(MIP(u,v,G^t))$} \label{alg:2:1}
    \STATE return; \label{alg:2:2}
\ENDIF \label{alg:2:3}
\FOR{each node $i$ with $prob(MIP(i,u,G^t))> \theta$} \label{alg:2:4}
    \FOR{each node $j$ with $prob(MIP(v,j,G^t))> \theta$} \label{alg:2:5}
        \IF {$prob(MIP(i,j,G^t))< \theta$ and \\ $prob(MIP(i,j,G^{t'}))> \theta$} \label{alg:2:6}
            \STATE $deltaInf[i] += prob(MIP(i,j,G^{t'}))\times (1-prob(j,S))$ \label{alg:2:7}
        \ENDIF \label{alg:2:8}
        \IF {$prob(MIP(i,j,G^t))> \theta$ and \\ $prob(MIP(i,j,G^{t'}))> \theta$} \label{alg:2:9}
            \STATE $deltaInf[i] += (prob(MIP(i,j,G^{t'}))-prob(MIP(i,j,G^{t})))\times (1-prob(j,S))$ \label{alg:2:10}
        \ENDIF \label{alg:2:11}
    \ENDFOR \label{alg:2:12}
\ENDFOR \label{alg:2:13}
\end{algorithmic}
\end{algorithm}

Consider the case when a new edge $e=(u,v,w)$ is introduced between two existing node $u$ and $v$. We denote the graph before and after such a topology change as ${G^t}$ and ${G^{t'}}$, and the current seed set is $S$. The detailed algorithm is described in algorithm \ref{alg:2}. According to the principle of localization \cite{chen10}, if the propagation probability $w$ is smaller than the specified threshold $\theta$, or not bigger than the probability of $MIP(u,v,G^t)$, edge $e$ can be simply neglected and there is no need to update any node's influence spread (lines \ref{alg:2:1}-\ref{alg:2:3}). Otherwise, the newly-added edge $e$ would become the $MIP(u,v,G^{t'})$. As a result, each node $i$ whose maximum influence path to $u$ has a influence probability larger than $\theta$ is likely to experience a rise in terms of influence spread (line \ref{alg:2:4}) because node $i$ may influence more nodes through the new edge $e$. So, we then check the probability of the maximum influence path from $i$ to $v$ and its successors in ${G^t}$ and ${G^{t'}}$. Based on the two probabilities, we divide the problem into two small cases:

The first case is when the probability of maximum influence path from $i$ to $j$ in ${G^t}$ is smaller than $\theta$ while that in ${G^{t'}}$ is larger than $\theta$ (lines \ref{alg:2:5}-\ref{alg:2:6}). Here $j$ denotes the node whose probability of $MIP(v,j,G^t)$ is larger than $\theta$. In such a case, node $i$ build a new path to $j$ through the new edge $e$ which increases the influence spread of $i$ by $prob(MIP(i,j,G^{t'}))\times(1-prob(j,S))$ (line \ref{alg:2:7}). Here $prob(j,S)$ is the probability of that node $j$ is influenced by the current seed set $S$, which is defined as follows:
\begin{equation*}\label{equ:3}
prob(j,S)=
\begin{cases}
1, & \text{if $j \in S$} \\
1-\prod_{w\in n(j)} {1-prob(w,S) \cdot  p(w,j)}, & \text{if $j \notin S$}
\end{cases}
\end{equation*}
Here $n(j)$ denotes the in-neighbour set of $j$.

The second case is when the probability of maximum influence path from $i$ to $j$ is larger than $\theta$ in both ${G^t}$ and ${G^{t'}}$ (lines \ref{alg:2:9}-\ref{alg:2:11}). In this case, the influence increase of node $i$ is $(prob(MIP(i,j,G^{t'}))-prob(MIP(i,j,G^{t})))\times(1-prob(j,S))$.

We treat the network dynamics from $G^t$ to $G^{t+1}$ as a finite change stream $c_1,c_2,\cdots,c_i,\cdots$ where each change $c_i$ is one of the six topology changes we described above. When all the changes in the change stream are processed, we can obtain the influence spread change for all the nodes.
\subsection{Potential Top-$K$ Influential Users Identification}
\label{sec:43}
Inspired by the observations of Section \ref{sec:3}, we design a pruning strategy to reduce the search space of potential influential nodes in this subsection. It is assumed that we only know who are the top-$K$ influential nodes in graph $G^t$, but their detailed influence spreads are beyond our knowledge. The reason are mainly twofold. First, several influence maximization algorithms, such as DegreeDiscount \cite{chen09} and SA \cite{sa}, do not calculate the influence spread information to identify influential users so that such information are unavailable. Second, even though these information are ready, storing them will cost as much as $O(nK)$ memory space where $n$ is the number of node in $G^t$. Since real-world social networks are typically of large scale, this will introduce serious storage overhead and directly affect the scalability.

From the preferential attachment rule, we know that the influence spread changes of those high-degree nodes should be much greater than the ordinary nodes. Moreover, according to the power-law distribution, such high-degree nodes only account for a small part of the whole nodes. Consequently we can pick out nodes only experiencing major increases or with high degrees because these nodes are of great potential to become the top-$K$ influential nodes in $G^{t+1}$. Then we only calculate the actual influence spread for these selected nodes while ignoring the others.
In this way, a large percent of nodes are pruned and the search space is largely narrow. It should be noted that a smart pruning strategy is of key importance since a poor selection might either affect the efficiency or reduce the accuracy in terms of influence spread. We describe the details of our pruning strategy as follows:

\begin{enumerate}
\item In the $i$th iteration, if the influence spread of the previous influential node $S_i^t$ increases in $G^{t+1}$, the chosen nodes are those with a larger influence spread change than $deltaInf[S_i^t]$; \label{item1}

In most cases, the influential nodes will attract a great number of new nodes and establish new links. Thus, their influence spreads will increase drastically. In such a case, the nodes whose influence spread changes are smaller than the influential nodes are completely impossible to become the most influential node in $G^{t+1}$. Therefore, when the influence spread of the previous influential nodes increase, we only select those whose influence spread changes are larger than the influential nodes in $G^t$. According to the preferential attachment rule, such a pruning method can greatly narrow the search space and reduce the amount of computation.
\item In the $i$th iteration, if the influence spread of the previous influential node $S_i^t$ decreases in $G^{t+1}$, in addition to qualification \ref{item1}, the nodes are further selected to hold a sufficiently large degree or experience a sufficiently great increase. In order to formally define ``large degree" and ``great increase", here we set an threshold $\eta$ to tradeoff between running time and influence spread. Here the nodes with sufficiently large degrees (or great increase) are defined as the set of node $v_j$ whose degree (or degree increase ratio) is among the top $\eta$ percent of all nodes in $G^{t+1}$. The degree increase ration of $v_j$ is defined as $degree_j^{t+1}/degree_j^{t}$ where $degree_j^{t}$ denotes the degree of node $v_j$ in graph $G^{t}$. Experimental results in Section \ref{sec:5} will demonstrate that 5\% may stand as a good tradeoff between running time and influence spread.
     \label{item2}

\begin{table*}
\centering
\caption{Summary information of the real-world social networks}
\label{tab:1}       
\begin{tabular}{c|r|r|r|r|r|r}
\hline
\multirow{2}*{Datasets} & \multicolumn{3}{c|}{Nodes} & \multicolumn{3}{c}{Edges} \\
\cline{2-7}
& Initial Number & Final Number & Growth & Initial Number & Final Number & Growth\\ \hline\hline
Facebook & 12,364 & 61,096 & 394\% & 73,912 & 905,665 &1125\%\\ \hline
NetHEPT & 5,802 & 29,555 & 409\% & 57,765 & 352,807 & 511\%\\ \hline
Flickr & 1,620,392 & 2,570,535 & 58.6\% & 17,034,807 & 33,140,018 & 94.5\% \\ \hline
\end{tabular}
\end{table*}

It should be noted that although the case the influence spread of a previous influential node decreases during the evolution rarely happens, we consider it here for completeness. In this case, except for qualification \ref{item1}, we further select nodes because the number of nodes satisfying qualification \ref{item1} is relatively large which lead to mass computation. While in reality, a node with small degree has only very low probability to become an influential node. In order to select only the most potential nodes, we refine the requirement and additionally select the nodes with large degree and large increase. Consequently, the search space is strictly circumscribed and the computational complexity is greatly reduced.
\end{enumerate}

\begin{algorithm}
\caption{IncInf}
\label{alg:3}
\begin{algorithmic}[1]
\REQUIRE ~~\
${G^t}$, ${S^t}$, and $G^{t+1}$.
\ENSURE ~~\
the top-$K$ influential nodes $S^{t+1}$ in $G^{t+1}$.
\STATE Initialize $S^{t+1}=\emptyset$; \label{alg:3:1}
\FOR{$i=1$ to $K$} \label{alg:3:2}
    \FOR{each topology change $c_j$ from $G^{t}$ to $G^{t+1}$} \label{alg:3:3}
        \STATE calculate the influence spread change $deltaInf[\cdot]$; \label{alg:3:4}
    \ENDFOR \label{alg:3:5}
    \STATE select a set of potential nodes as $pn$ according to pruning strategy; \label{alg:3:6}
    \FOR{each node $v_l\in pn$}  \label{alg:3:7}
        \STATE calculate the marginal influence spread $\sigma _{S^{t+1}} (v_j)$; \label{alg:3:8}
    \ENDFOR \label{alg:3:9}
    \STATE select $v_{max}=\arg\max _ {v_j\in pn} {(\sigma _{S^{t+1}} (v_j))}$; \label{alg:3:10}
    \STATE $S = S \cup v_{max}$; \label{alg:3:11}
\ENDFOR \label{alg:3:12}
\end{algorithmic}
\end{algorithm}

After the potential nodes are selected, we calculate the actual influence spread of these nodes in $G^{t+1}$ and select the one with the maximum influence spread in each iteration. Algorithm \ref{alg:3} outlines the design of our proposed algorithm IncInf. IncInf iterates for $K$ round (line \ref{alg:3:2}) and in each round select one node providing the maximum marginal influence spread. Lines \ref{alg:3:3} - \ref{alg:3:5} calculate the influence spread change of each node caused by the topology evolution. Nodes with great potential to become top-$K$ influential are selected (line \ref{alg:3:6}) and their influence spread are computed in $G^{t+1}$ (lines \ref{alg:3:7} - \ref{alg:3:9}). And then the node providing the maximal marginal gain will be selected and added to the set $S^{t+1}$ (lines \ref{alg:3:10} - \ref{alg:3:11}).

\section{Experiments}
\label{sec:5}
In this section, we present the experimental results of our algorithm on identifying top-$K$ influential nodes in dynamic social networks. We examine two metrics, running time and influence spread, for evaluating the effectiveness as well as the execution efficiency of different algorithms. The experimental results are detailed in Section \ref{sec:52}, \ref{sec:53} and \ref{sec:54}.
\subsection{Experimental Setup}
\label{sec:51}

We choose three real-world social networks including Facebook social network, NetHEPT citation network, and Flickr social network. Table \ref{tab:1} summarizes the statistical information of the datasets.

\begin{itemize}
\item \textbf{Facebook.} This dataset is the friendship relationship network among New Orleans regional network on Facebook, spanning from Sep 2006 to Jan 2009 \cite{facebook}. There are more than $60K$ users connected together by as much as $1.5M$ links in the social network. 41.4\% of these edges contain no time information and are thus discarded. In our experiments, the nodes and links from Sep. 2006 to Apr. 2007 are used as the first snapshot and then network snapshots are recorded every 3 months.
\item \textbf{NetHEPT.} This is an academic citation network \cite{nethept} extracted from ``High Energy Physics-Theory" section of the arXiv over the period from 1992 to 2003, and covers the citations within a dataset of $28K$ papers with $352K$ edges. In our experiments, the citation links of the first three year (i.e. from 1992 to 1994) are considered as the basic graph and the network snapshots are recorded once a year.
\item \textbf{Flickr.} This dataset \cite{flickr} contains the user-to-user links crawled from the Flickr social network daily over the period from Nov. 2, 2006 to Dec. 3, 2006 and again from Feb. 3, 2007 and May 18, 2007, representing a total of 104 days of growth. There are totally $2.5M$ Flickr users and $33M$ links. During this period of observation, over 9.7 million new links are formed and over 950,000 new users joined the network. In our experiments, we use the network before Nov. 2, 2006 as the basic graph and another five snapshots are recorded on Dec. 3, Feb. 3, Mar. 3, Apr. 3, and May 18.
\end{itemize}

We compare our algorithm with four static algorithms: \textbf{MixGreedy}, \textbf{ESMCE}, \textbf{MIA} and \textbf{Random}. MixGreedy is an improved greedy algorithm on the IC model proposed by Chen et al. in \cite{chen09}. ESMCE is a power-law exponent supervised estimation approach designed by Liu et al. in \cite{esmce}. MIA is a heuristic that uses local arborescence structures of each node to approximate the influence propagation \cite{chen10}. Random is a basic heuristic that randomly selects $K$ nodes from the whole datasets.

The propagation probability of the IC model is selected randomly from 0.1, 0.01, and 0.001 for each network snapshot, and we run simulations on networks 10000 times and take the average of the influence spread.
\subsection{Efficiency Study}
\label{sec:52}
In this subsection, the efficiency of our proposed algorithm is studied and compared with corresponding static algorithms, MixGreedy and MIA, through experiments on the Facebook, NetHEPT and Flickr datasets. The experiments are conducted on a PC with Intel Core i7 920 CPU @2.67 GHz and 6 GB RAM. The running time of four algorithms are measured by selecting 50 seeds from the whole dataset.
\begin{figure*}
\centering
  \includegraphics[width=160mm]{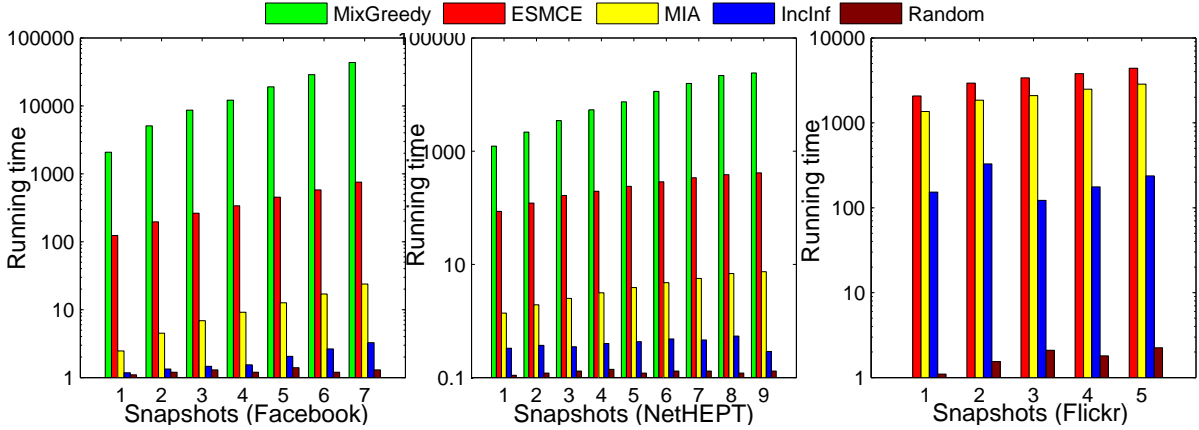}
\caption{The time costs of different algorithms on three real-world datasets.}
\label{fig:6}       
\end{figure*}

The time costs of different algorithms are illustrated in Figure \ref{fig:6} where we record the total time cost for each snapshot of the three datasets. Since incremental and static algorithms have the same time cost in the initial snapshot, thus they are omitted in the figure. The experimental results show that the time costs of our algorithm on each snapshot are obviously less than those of static algorithms. Obviously, MixGreedy takes the longest time among four kinds of influence maximization algorithms. It takes MixGreedy more than as much as 6 hours to identify the top 50 influential nodes on the final NetHEPT dataset, while the time is even longer on the larger dataset Facebook. Moreover, MixGreedy is not feasible to run on the largest dataset Flickr due to the unbearably long running time. ESMCE, benefiting from its sampling estimation method, runs much faster than MixGreedy, but it still takes as much as 3511 seconds on average to run on the five snapshots of Flickr. Compared with two greedy algorithms, the heuristic MIA performs much better. It only takes MIA 23.8 seconds to run on the final Facebook graph. When running on the Flickr dataset with as much as $2.5M$ nodes and $33M$ edges, however, its speedup is far from satisfactory, since it still needs more than 45 minutes to finish.
While our proposed algorithm, IncInf, outperforms all the static algorithms in terms of efficiency. In particular, IncInf is almost four orders of magnitude faster than the MixGreedy algorithm on the Facebook dataset. While compared with the MIA heuristic, the speedup of IncInf is $8.41\times$ and $6.94\times$ on the Facebook and NetHEPT datasets, respectively; What's more, when applied on the largest dataset Flickr, IncInf can achieve as much as $20.65\times$ speedup on average.
This is because IncInf only computes the incremental influence spread changes and adaptively identifies the influential nodes based on the previous influential nodes and the current influence spread changes. The experimental results clearly validate the efficiency advantage of our incremental algorithm IncInf.
We can also observe that the running time of IncInf is not monotone like other algorithms as the time evolves. This is because the running time of IncInf is closely related to the topology change between two graph snapshots. An evident change in topology will usually lead to a relatively long running time and vice versa.
Without doubt, Random runs the fast among all the algorithms. However, as we will show in Section \ref{sec:53}, its accuracy is much worse and unacceptable when developing real-world viral marketing strategies.

\begin{figure}
\centering
  \includegraphics[width=60mm]{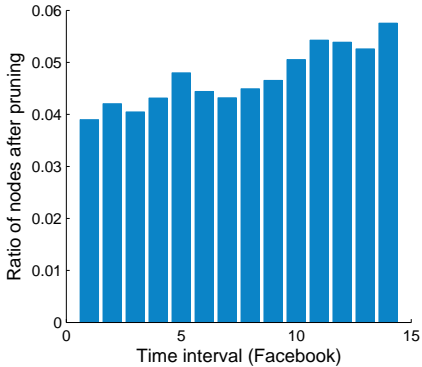}
\caption{The effect of pruning strategy on the Facebook dataset.}
\label{fig:7}       
\end{figure}
We also test the effect of our pruning strategy. Here we take the Facebook dataset as an example; the results on other datasets are similar and thus omitted. Different from other experiments, we  recorded the Facebook graph from Sep. 2006 to Oct. 2007 (14 months) as snapshot $A$ in this experiment. After that we take snapshots every month as snapshot $B$. We use IncInf to find the top-$K$ influential nodes in snapshot $B$ based on ones in snapshot $A$. The result is shown in Figure \ref{fig:7}. The x axis is the time interval between snapshot $A$ and $B$, and the y axis the ratio of the number of nodes after pruning to the total number of nodes in snapshot $B$. The minimum and maximum pruning ratios are 3.90\% and 5.86\% respectively, with a mean ratio of 4.72\% on all the 14 time intervals between snapshot $A$ and $B$. This demonstrates that our pruning strategy can effectively limit the search space into a small percent of nodes. We can also see in Figure \ref{fig:7} that with the increase of time interval, the ratio, although not monotone, generally becomes larger. This is mainly because a longer time interval means a larger amount of topology changes, and basically more nodes will be potential to become influential nodes.
\subsection{Effectiveness Study}
\label{sec:53}
\begin{figure*}
\centering
  \includegraphics[width=160mm]{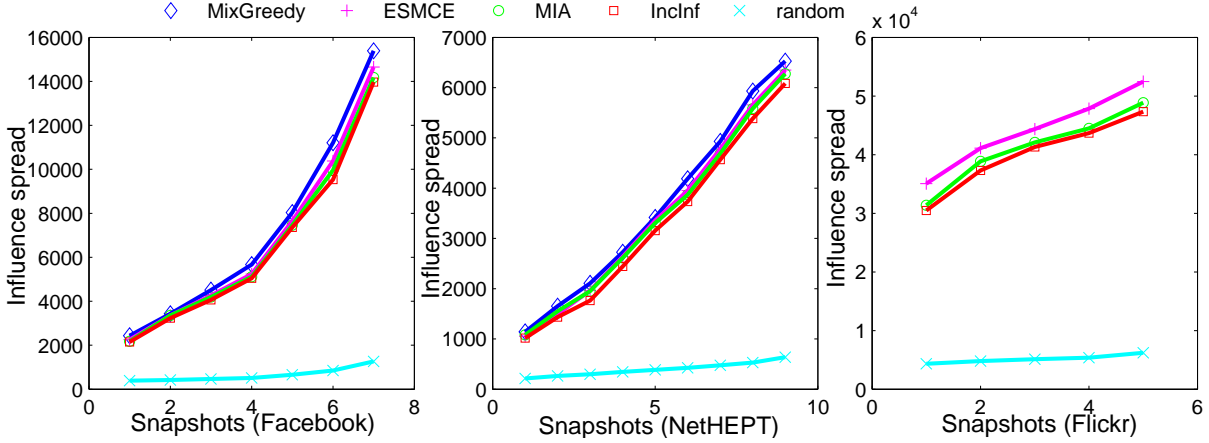}
\caption{The influence spread of different algorithms on three datasets.}
\label{fig:8}       
\end{figure*}
In this subsection, we study the influence spread of the top-$K$ influential nodes selected by our algorithm as well as other static algorithms. The influence spreads of different algorithms are measured as the number of nodes that are influenced by the top-50 influential nodes selected. Obviously, the higher the influence spread, the better the effectiveness. We have not test the performance of MixGreedy on the Flickr dataset as the running time is excessively long.

Figure \ref{fig:8} shows the experimental results. MixGreedy outperforms all the other algorithms in terms of influence spread. However, the efficiency issue limits its application to large-scale dataset such as Flickr. The performance of ESMCE, MIA and IncInf almost match MixGreedy on the Facebook dataset, while on NetHEPT, the gaps become larger but remain acceptable (only 3.4\%, 4.7\% and 5.1\% lower than MixGreedy on average). When applied to the Flickr dataset, ESMCE performs the best since ESMCE strictly control the error threshold by iterative sampling. Compared with MIA, IncInf shows very close performance and is only 2.87\% lower on average of all five snapshots, which demonstrates the effectiveness of our proposal. Random, as the baseline heuristic, clearly performs the worst on all the graphs. Actually, the influence spread of Random is only 15.6\%, 12.1\% and 10.9\% of that of IncInf on Facebook, NetHEPT and Flickr, respectively.

We shall note that the reason IncInf has slightly lower influence spread is mainly twofold. First, IncInf restrict the influence into local regions to speed up the
computation of influence spread changes, which will affect the effectiveness. Second, a pruning strategy is designed to narrow down the search space based on the influence spread changes and previous top-$K$ information. Despite slight loss in effectiveness, as aforementioned, the disparity is small and acceptable. More importantly, IncInf gains remarkable improvement in efficiency.

\subsection{Tuning of Parameter $\theta$ and $\eta$}
\label{sec:54}
First, we study how effectively the localization parameter $\theta$ of IncInf represents a tradeoff between efficiency and effectiveness. We run IncInf with different values of $\theta$ on the final Facebook and NetHEPT graphs. The running time and influence spread are measured based on seed size $K=50$.

\begin{figure*}
\centering
\subfloat[Facebook]{
\includegraphics[width=70mm]{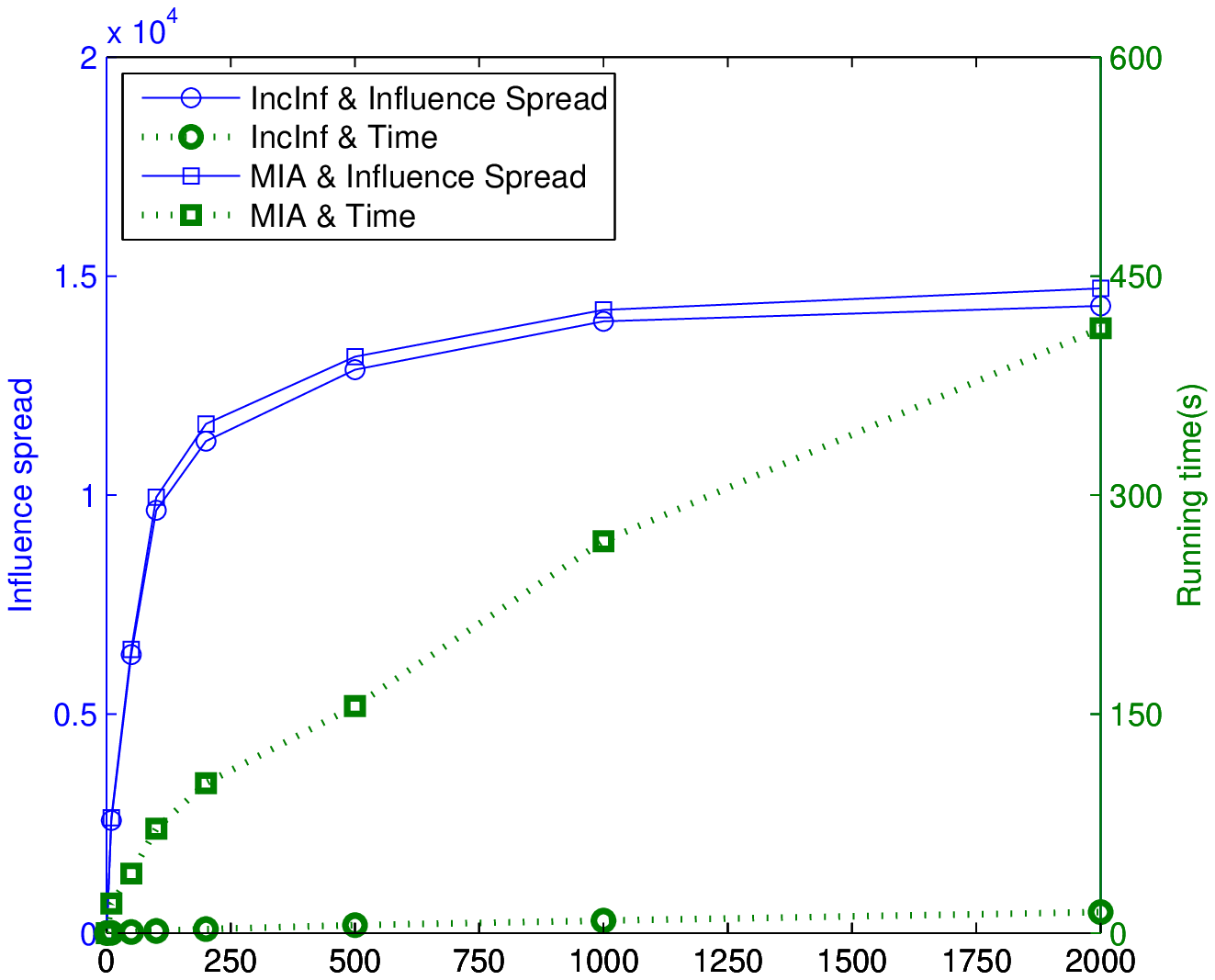}
\label{fig:8a}
}
\hfil
\subfloat[NetHEPT]{
\includegraphics[width=70mm]{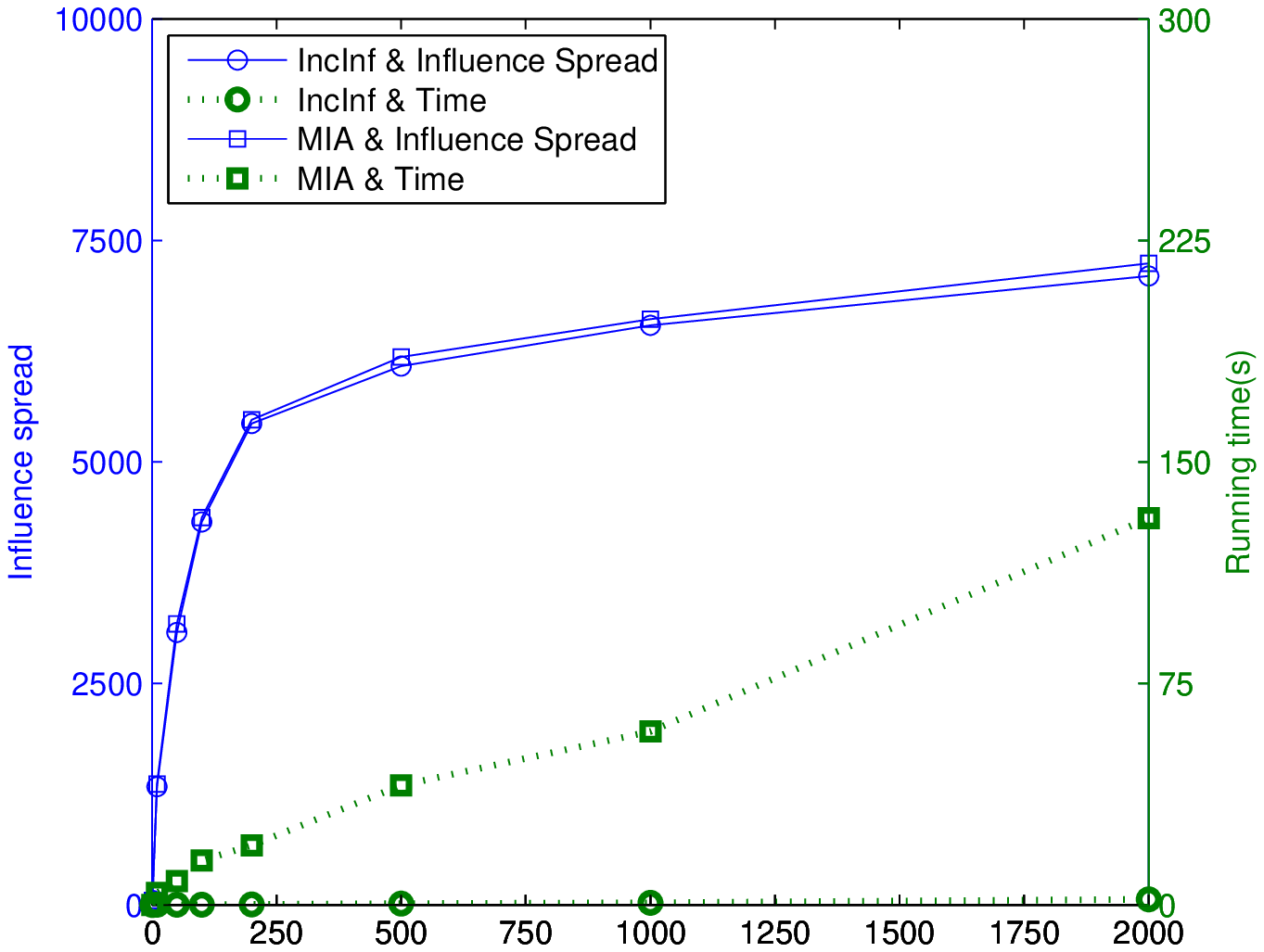}
\label{fig:8b}
}
\caption{The effect of tuning of $\theta$ on running time and influence spread.}
\label{fig:88}
\end{figure*}

The experimental results are shown in Figure \ref{fig:88}. Note that the x axis represents the reciprocal of $\theta$. We observe that $\theta$ acts as a tradeoff between efficiency and effectiveness: with the decrease of $\theta$, IncInf and MIA achieve better influence spread. However, this is gain at the cost of longer running time, i.e., poor efficiency. For example, when we reduce $\theta$ from 1/200 to 1/500 on the Facebook dataset, the influence spread of IncInf increases by 15.4\% while the running time is $1.12\times$ longer. Moreover, we can observe that the influence spread of IncInf almost match that of MIA in all values of $\theta$. For example, IncInf is only 1.87\% lower than MIA in influence spread when $\theta$ is set to 1/200 in the NetHEPT dataset. But IncInf shows overwhelming advantages in terms of running time. When $\theta$ is set to 1/500 in Facebook, IncInf needs only 5 second to identify the top-50 influential nodes while it takes MIA more than 150 second to finish the same work. More importantly with the decrease of $\theta$, the influence spread increases sharply at the beginning but the increase is no longer that significant after $\theta$ is lowered to a certain level. On the contrary, the running time is almost linear to $1/\theta$. This suggests that the knee point of the influence spread curve can serve as a good tuning point of $\theta$ where we could obtain the best gain from both influence spread and running time.

\begin{figure}
\centering
  \includegraphics[width=70mm]{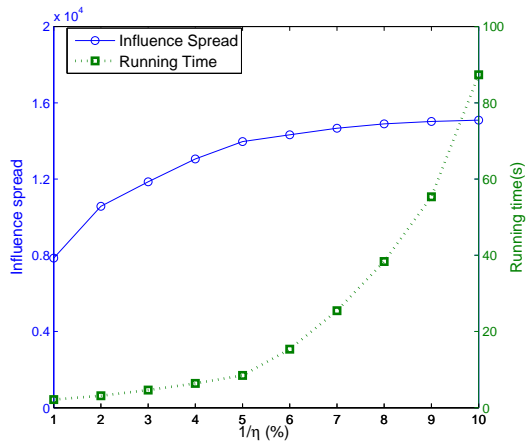}
\caption{The effect of tuning of $\eta$ on running time and influence spread.}
\label{fig:10}       
\end{figure}

Then, we will evaluate the sensitivity of pruning threshold $\eta$ in terms of influence spread and running time. The results are illustrated in Figure \ref{fig:10}.
From figure \ref{fig:10} we can see that, with the increase of $\eta$, the running time increase gently at the beginning and then turns into a sharp boost. For example, when we increase $\eta$ from 1\% to 5\%, the running time of IncInf on the Facebook dataset only increase from 2.13s to 8.47s, while it dramatically increases from 8.47s to 87.35 when $\eta$ is tuned from 5\% to 10\%. This phenomenon is closely related to the power-law distribution of degree in social network; when $\eta$ set large, a relatively large number of potential nodes would be selected.

In terms of influence spread, as the increase of $\eta$, more nodes are selected as potential nodes which will guarantee better influence spread. Different from the running time, the influence spread grows rather rapid at the beginning, and then gradually slows down. The influence spread on the Facebook Dataset is 7854 when $\eta$ is set to 1\%, and rapidly grow to 13967 when the maximum error threshold is 5\%. After that, the growth trend slows down and the influence spread is about 15091 as $\eta$ increases to 10\%. This reason to explain such phenomenon is that the top-$K$ influential nodes are mainly selected from high degree nodes. Therefore, when $\eta$ becomes larger, although more nodes would be selected, their contribution to influence spread are relatively small, thus the growth trend slows down.
Based on the above observation, here we suggest that 5\% may stand as a good tradeoff between running time and influence spread.

\subsection{Discussions}
Experimental results demonstrate that our proposed IncInf algorithm significantly reduces the execution time of state-of-the-art static influence maximization algorithm while maintaining satisfying accuracy in terms of influence spread. Although IncInf  performs better, it has a few limitations for further improvement.

First, IncInf directly depends on previous information of top-$K$ influential nodes for effective pruning, while sometimes such information are incomplete, or even unavailable. We plan to study this problem later. Second, IncInf is designed for the IC model which may somehow limit its application. But we believe our idea of incremental computation for influence maximization could be properly extended to other influence diffusion models.

\section{Related Work}
\label{sec:6}
Influence maximization on static networks has attracted a lot of attentions. The hill-climbing greedy algorithms proposed by Chen et al. suffers from low efficiency, and many efficient algorithms have been proposed recently to address this problem. Leskovec et al. \cite{jure07} exploit the submodularity of influence spread function and develop an optimized greedy algorithm, CELF, which is much faster than basic greedy algorithm. Chen et al. \cite{chen09} propose MixGreedy which computes the influence spread for each seed set in one single simulation and incorporates the CELF optimization. MIA \cite{chen10} uses local arborescence structures of each node to approximate the influence spread, thereby gaining efficiency by restricting computations and updates only on the local regions. However, MIA only considers static networks while in this paper we specifically design an incremental algorithm for evolving social networks. Recently, Wang et al. \cite{cga} propose a Community Greedy Algorithm (CGA) that took community property into account.
Goyal et al. propose CELF++\cite{celf++} further exploits the property of submodularity of the spread function to avoid unnecessary re-computations of marginal gains, and considerably improves the efficiency of CELF algorithm.
IRIE \cite{irie} is also a heuristic proposed by Jung et al. that incorporates influence ranking algorithm with influence estimation method to achieve scalability.
Chen et al. \cite{batch} propose a BatchGreedy algorithm for active learning and demonstrated through experiments that BatchGreedy could considerably improved the effectiveness of previous greedy algorithms.
Liu et al. \cite{liutpds2014} design a new framework to accelerate the influence maximization by leveraging the parallel processing capability of GPU.
In \cite{leewww2014}, Lee et al. propose GIS with a similar idea of influence localization, but they didn't consider the dynamic feature of online social networks.
Cheng et al. \cite{sigir2014} present IMRank to solve the IM problem via finding a self-consistent ranking.

The influence maximization problem on dynamic social networks still remains largely unexplored to date. Habiba et al. \cite{habiba2007} propose a dynamic social network model which is different from ours. In their proposal, the network keeps evolving during the process of influence propagation, and their goal is to find the top-$K$ influential nodes over such a dynamic network. When compared to \cite{habiba2007}, our work is based on snapshot graph model and our goal is to incrementally identify top-$K$ influential nodes based on the topology changes of two adjacent snapshots. Chen et al. \cite{chentime2012} extend the IC model to incorporate the time delay aspect of influence diffusion among individuals in social networks, and consider time-critical influence maximization, in which one wants to maximize influence spread within a given deadline. While in \cite{gomez2012}, the authors consider a continuous time formulation of the influence maximization problem in which information or influence can spread at different rates across different edges. Charu Aggarwal et al. \cite{aggarwal2012} try to discover influential nodes in dynamic social networks and they design a stochastic approach to determine the information flow authorities with the use of a globally forward approach and a locally backward approach. Their influence model and target are different from ours.
Zhuang et al. \cite{zhuang2014} argue that the evolution of online social network could not be fully observed and focus on the problem of designing a proper probing strategy so that the actual influence diffusion process can be best uncovered with the probing nodes.


\section{Conclusion and Future Work}
\label{sec:7}
In this paper, we consider the influence maximization problem in evolving social networks, and propose an incremental algorithm, IncInf, to efficiently identify top-$K$ influential nodes in dynamic social networks. Taking advantage of the structural evolution of networks and previous information on individual nodes, IncInf substantially reduces the search space and adaptively selects influential nodes in an incremental way. Extensive experiments demonstrate that IncInf significantly reduces the execution time of state-of-the-art static influence maximization algorithm while maintaining satisfying accuracy in terms of influence spread.

There are several future directions for this research. First, IncInf has large potential to fit into modern parallel computing framework. This is because IncInf restricts the computation of influence spread changes into local regions, which could ease the partition of social graph for parallel computation. Moreover, the proposed pruning strategy could be effectively performed in parallel. Second, our current IncInf algorithm is derived from the basic IC model. We believe the conception of incremental computation for influence maximization could be properly extended to other influence diffusion model, such as another classic LT model. Third, although there have been a few research \cite{tang2009,cha2010} about how to measure the propagation probability, however this problem is not yet well addressed especially for large-scale dynamic social networks.

%

%
%


\bibliographystyle{abbrvnat}


\end{document}